\documentstyle[11pt,newpasp,twoside,epsf]{article}
\begin{document}
\title{The Millennium Outburst of the Cool Hypergiant $\rho$~Cassiopeiae : Spectroscopy and Modeling}

\author{A. Lobel\altaffilmark{1}, A. K. Dupree, R. P. Stefanik, G. Torres} 
\affil{Harvard-Smithsonian Center for Astrophysics, 60 Garden Street, Cambridge MA 02138} 

\author{G. Israelian}
\affil{Instituto de Astrofisica de Canarias, E-38200 La Laguna, Tenerife, Spain} 

\author{N. Morrison}
\affil{Ritter Astrophysical Research Center, University of Toledo, Toledo, OH 43606} 

\author{C. de Jager, H. Nieuwenhuijzen}
\affil{SRON Laboratory for Space Research, Sorbonnelaan 2, 3584 CA Utrecht, the Netherlands } 

\author{I. Ilyin}
\affil{Astronomy Division, PO Box 3000, 90014 University of Oulu, Finland} 

\author{F. Musaev}
\affil{Special Astrophysical Observatory, Nizhnij Arkhyz 369167, Russia}

\altaffiltext{1}{Guest investigator of the UK Astronomy Data Centre.}

\begin{abstract}

Between 2000 June and September an exceptional variability phase occurred 
in the peculiar F-type hypergiant $\rho$ Cas,
when the $V$-brightness dimmed by at least a full magnitude.  The star recovered 
from this deep minimum by 2001 April. It is the third outburst of $\rho$ Cas 
on record in the last century. We observe TiO absorption bands in 
high-resolution optical and near-IR spectra obtained with the Utrecht Echelle Spectrograph
during the summer of 2000. TiO formation in the outer atmosphere occurred 
before the deep brightness minimum. Atmospheric models reveal that the 
effective temperature decreases by at least 3000 K, and the TiO shell is 
driven supersonically with $\dot{M}$$\simeq$5.4$\times$$10^{-2}$ $M_{\sun}\,yr^{-1}$. 
Strong episodic mass loss and TiO have also been observed during the 
outbursts of 1945-47 and 1985-86. 

An analysis of the exceptional outburst spectra of 2000--01 is provided, 
by comparing with high-resolution optical spectra of the early M-type 
supergiants $\mu$ Cep ($\rm Ia$) and Betelgeuse ($\rm Iab$). 
During the outburst, central emission appears above the local continuum level 
in the split Na $D$ lines. A prominent optical emission line spectrum 
appears in variability phases of fast wind expansion. The radial velocity 
curves of H$\alpha$, and of photospheric metal absorption lines signal a 
very extended, and velocity stratified dynamic atmosphere. The outburst 
spectra indicate the formation of a low-temperature, optically thick 
circumstellar gas shell of 3$\times$$10^{-2}$ $M_{\odot}$ during ~200 d, 
caused by dynamic instability of the upper atmosphere of this pulsating 
massive supergiant near the Eddington luminosity limit. 

We present an equation that correctly predicts the outburst time-scale,
whereby the shell ejection is driven by the release 
of hydrogen recombination energy. The observations reveal that during the outburst 
the atmospheric hydrogen recombination zone dissipates $\sim$5~\% of the 
stellar luminosity into the circumstellar environment, by driving a superwind.
We observe that the mass-loss rate during the outburst is of 
the same order of magnitude as has been proposed for the outbursts of $\eta$ Carinae. 
These phases of punctuated mass-loss represent the major 
mass-loss mechanism of cool massive hypergiants.
The research results in this paper are described in further detail in Lobel et al. (2003). 
A spectral movie sequence of the outburst is available at   
http://cfa-www.harvard.edu/$\sim$alobel.    
\end{abstract}


\index{*$\rho$ Cas}
\index{*Betelgeuse| $\alpha$ Ori}
\index{*$\mu$ Cep}

\section{Introduction}

Cool hypergiant stars as $\rho$ Cas (HD 224014) and HR~8752 are thought to be post-red supergiants, 
rapidly evolving toward the blue supergiant phase (de Jager 1998).
They are rare enigmatic objects, which we are continuously monitoring with 
high spectral resolution since about one decade.
Yellow hypergiants ($\rm Ia^{+}$) are very important objects to investigate the 
physical causes for the luminosity limit of evolved stars.
A spectroscopic study during 1993-95 showed that $\rho$ Cas is a slowly 
pulsating supergiant with changes in the optical spectrum corresponding to 
variations in $T_{\rm eff}$ of less than 750 K ($T_{\rm eff}$=7250 K $-$ 6500 K; 
Lobel et al. 1998).

During a famous outburst of $\rho$ Cas in 1945-47 many zero-volt excitation energy lines appeared, 
not previously observed in its absorption spectrum. These atomic lines, 
normally observed in M-type supergiants, were strongly blue-shifted, signaling the ejection 
of a cool circumstellar gas shell (Beardsley 1961). In the following two decades after 
this event, several papers were published in the astrophysical literature, 
discussing $\rho$ Cas' peculiar spectroscopic changes in the years during and after 
this dramatic outburst (e.g. Popper 1947;  Thackeray 1948). 
Between 1945 and 1946 the star rapidly dimmed and the spectrum developed optical and near-IR TiO bands. 
Within a couple of years, the hypergiant brightened up by nearly a full magnitude, 
and a mid G-type spectrum was recovered around 1950. More recently, 
Boyarchuk, Boyarchuk, \& Petrov (1988) also reported the detection of TiO bands 
in the optical spectrum during a more moderate outburst of 1985-87. 

This paper discusses the spectroscopic changes we monitored during the 
2000-01 outburst. We analyze prominent TiO bands newly detected in the spectra, 
before and during the outburst minimum. A brief discussion of the stellar outburst 
physics is also provided.

\section{Radial Velocity and Brightness Curves}

The upper panel of Figure~1 shows photo-electric observations of $\rho$ Cas 
in the $V$-band ({\it black dots}) by Percy, Kolin, \& Henry (2000) over the past decade, 
supplemented with visual magnitude estimates from $AAVSO$, and from $AFOEV$
(French Association of Variable Star Observers) during the outburst 
of late 2000 ({\it cyan dots}). The brightness curve shows semi-regular 
variability, with the deep outburst minimum of $V$$\sim$$5^{\rm m}.3$ in
2000 September--November, preceded by a conspicuously bright visual maximum ($V$$\sim$$4^{\rm m}$) 
in 2000 March. 

\begin{figure}
\plotfiddle{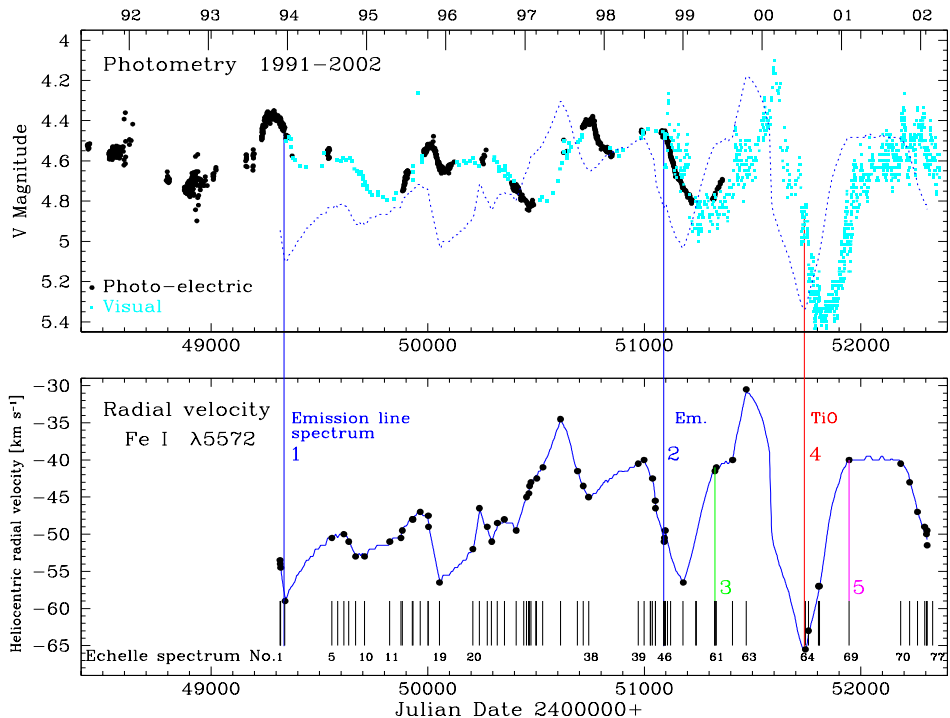}{5.0cm}{0}{54}{50}{-198}{-140}
\vspace*{4cm}
\caption {The $V$-brightness curve of $\rho$ Cas is compared in the upper panel to the radial velocity curve,
observed over the past 8.5 years. Observation dates of echelle spectra are marked with short vertical lines in 
the lower panel. The radial velocity curve of Fe~{\sc i} $\lambda$5572 
shows a strong increase of the photospheric pulsation amplitude before the outburst of fall 2000 
(JD 2451800$-$JD 2451900), when TiO bands develop ({\em marked TiO}). 
} \label{fig-1}
\end{figure}

The lower panel shows the radial velocity curve which has been monitored from the 
unblended Fe~{\sc i} $\lambda$5572 absorption line ({\it black dots}).
The radial velocity curve, determined from a linear interpolation of the 
temporal Fe~{\sc i} line profile changes, is compared with the $V$-magnitude 
curve in the upper panel ({\it blue dotted line}). We observe that the star becomes 
brightest for variability phases when the atmosphere rapidly expands. 
$V_{\rm rad}$ decreases by $\sim$20 $\rm km\,s^{-1}$ in less than 200 d during 
the outburst event. The short black vertical lines mark a total of 78
echelle spectra observed with high-resolution spectrographs
over the past 8.5 years.

The spectra have been obtained from our long-term monitoring campaigns 
with four telescopes in the northern hemisphere;  the Utrecht Echelle Spectrograph 
of the William Herschel Telescope (La Palma, Canary Islands), the Sofin spectrograph
of the Nordic Optical Telescope (La Palma, Canary Islands), the Ritter Observatory
telescope (OH, USA), and the Zeiss-1000 telescope of the Special Astrophysical Observatory 
of the Russian Academy of Science (Zelenchuk, Russia). 
The spectra marked with a colored long vertical line ({\it labeled 1 to 5}) in Figure~1 
are discussed below.  

\begin{figure}
\plotfiddle{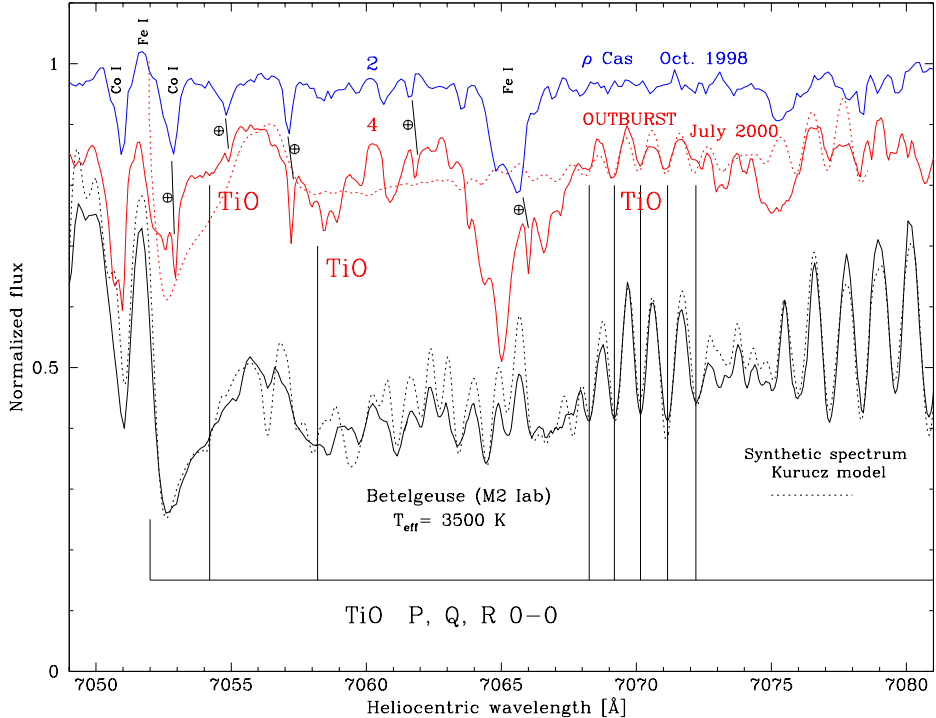}{5.0cm}{0}{50}{50}{-170}{-140}
\vspace*{4cm}
\caption {Near-IR TiO absorption 
bands are observed during the outburst of $\rho$ Cas in 2000 July.
The weak TiO bands around 7070~\AA\, have distinct shapes,
also observed in Betelgeuse. The synthetic 
spectrum of $\rho$ Cas is computed with only TiO lines, 
while that of Betelgeuse also includes molecular and atomic lines. 
} \label{fig-2}
\end{figure}

\section{Formation of Titanium-oxide Bands during the Outburst}

We observe the formation of titanium-oxide (TiO) absorption $\gamma$-system bands 
in the (WHT-UES) outburst spectra  of $\rho$~Cas in July 2000 ({\it red solid line in Fig. 2}). 
The calculated spectrum with only TiO lines is shown in Figure 2 with the red dotted line. 
A best fit is obtained for a Kurucz model atmosphere of $T_{\rm eff}$=3750 K 
and $\log{g}$=0. The characteristic TiO bands around 7070~\AA\, ({\it marked with vertical lines})
are also observed in the supergiant Betelgeuse (M2 Iab) ({\it black solid line}),
computed with $T_{\rm eff}$=3500 K and $\log{g}$=$-$0.5 ({\it black dotted line})
(see Lobel \& Dupree 2000). 

The spectrum synthesis typically contains $\sim$1500 TiO lines per \AA\, from the P, Q, and 
R-branches, for five $ ^Z$Ti\,$\rm ^{16}O$ isotopomers ($Z$=46--50), 
with Earth abundance fractions (Schwenke 1998). The TiO bands in $\rho$ Cas signal the 
formation of a cool, optically thick, circumstellar gas shell with $T$$<$4000 K,
caused by the supersonic expansion of the outer atmosphere during the outburst.
We observe an expansion velocity of $v_{\rm exp}$=35$\pm$2~$\rm km\,s^{-1}$  
for these TiO bands, about 15--20 $\rm km\,s^{-1}$
faster than the atmospheric expansion velocity we determine from the atomic photospheric 
absorption lines. 

\begin{figure}
\plotfiddle{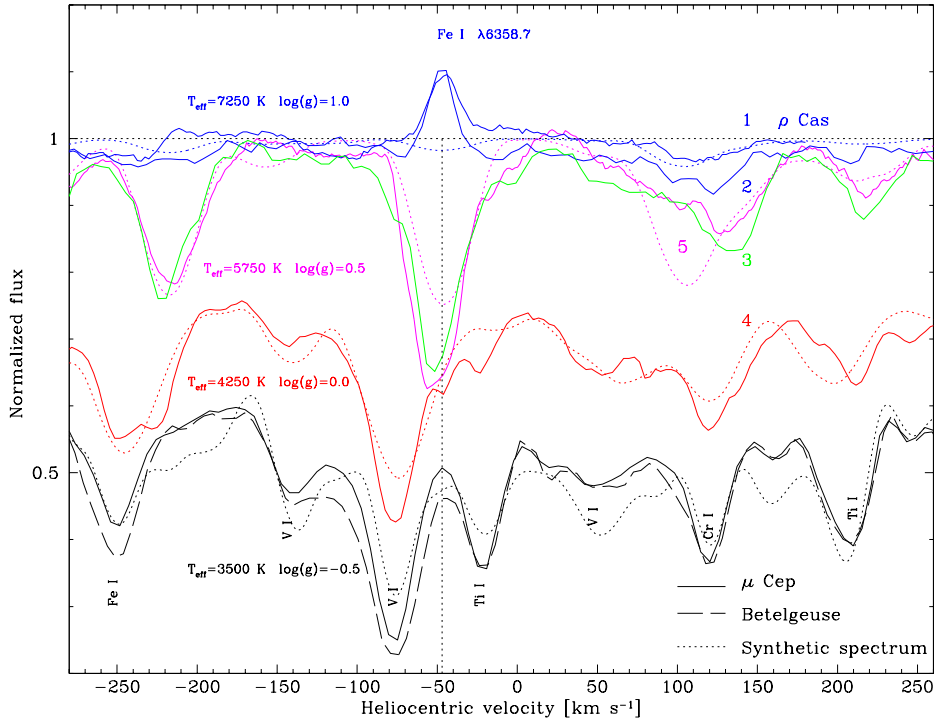}{5.0cm}{0}{50}{50}{-170}{-140}
\vspace*{4cm}
\caption {High-resolution spectra of 
$\rho$ Cas around Fe~{\sc i} $\lambda$6358 show that the photospheric spectrum strongly 
blue-shifts during the outburst. Synthetic spectrum calculations
indicate that $T_{\rm eff}$ decreases to $\simeq$4250~K, and the spectrum becomes comparable to the 
early M-type supergiants $\mu$~Cep  and Betelgeuse,
with $T_{\rm eff}$=3500~K. A prominent emission line spectrum is observed in $\rho$ Cas 
during phases of high $T_{\rm eff}$=7250~K, when the atmosphere rapidly accelerates outward.
} \label{fig-3}
\end{figure}

\section{Modeling the Outburst Spectrum}

In Figure 3 a best fit to the atomic spectrum, observed during the outburst of $\rho$ Cas
({\it solid red line labeled 4}),
is computed with $T_{\rm eff}$=4250~K and $\log{g}$=0 ({\it red dotted line}). 
The graph is centered in the heliocentric velocity scale around the Fe~{\sc i} 
and V~{\sc i} blend at $\lambda$6358. 
The vertical dotted line is drawn at the stellar rest velocity of 
$-$47 $\rm km\,s^{-1}$ (Lobel 1997). 
During the outburst, the entire photospheric spectrum Doppler shifts toward the shorter 
wavelengths, and develops many atomic absorption features also observed in Betelgeuse and $\mu$~Cep
(compare with lower black lines). 

\begin{figure}
\plotfiddle{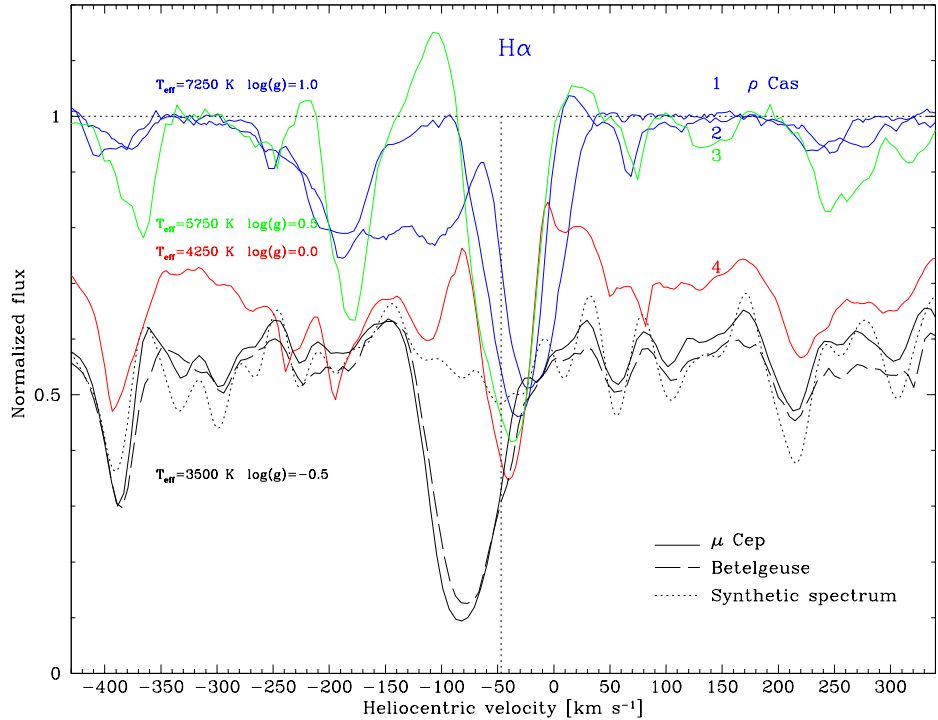}{5.0cm}{0}{50}{50}{-170}{-140}
\vspace*{4cm}
\caption {The H$\alpha$ absorption core becomes very weak during the outburst.
The line core does not blue-shift, and develops strong emission line 
wings. The line becomes filled in by recombination emission with the cooling 
of the H$\alpha$-envelope during the outburst. 
Strong emission is also observed in the blue wing of H$\alpha$
for the pre-outburst cycle, during the atmospheric collapse that precedes the large
$V$-brightness maximum before outburst. The H$\alpha$ line core of $\rho$ Cas is much weaker than in 
Betelgeuse and $\mu$ Cep, where it is excited by a permanent chromosphere.
} \label{fig-4}
\end{figure}

We also observe many peculiar metal emission lines appearing above 
the stellar continuum level during two phases with very fast atmospheric 
expansion in 1993 December ({\it blue vertical line labeled 1 in Figure 1}) and 1998 October ({\it labeled 2}). 
The accelerated stellar wind collides with circumstellar or interstellar material, 
which excites the permitted emission line spectrum. Further analyses of the formation of the 
peculiar emission line spectrum in $\rho$ Cas are presented in Lobel (1997 Chapter 4; 2001a).
For these variability phases we compute that $T_{\rm eff}$=7250~K ({\it upper blue dotted curve in Figure 3}).
The spectrum modeling therefore reveals a decrease in $T_{\rm eff}$ of at least 3000 K 
during the outburst of $\rho$ Cas in late 2000.

\section{Mass-loss Rate During the Outburst}

\begin{figure}
\plotfiddle{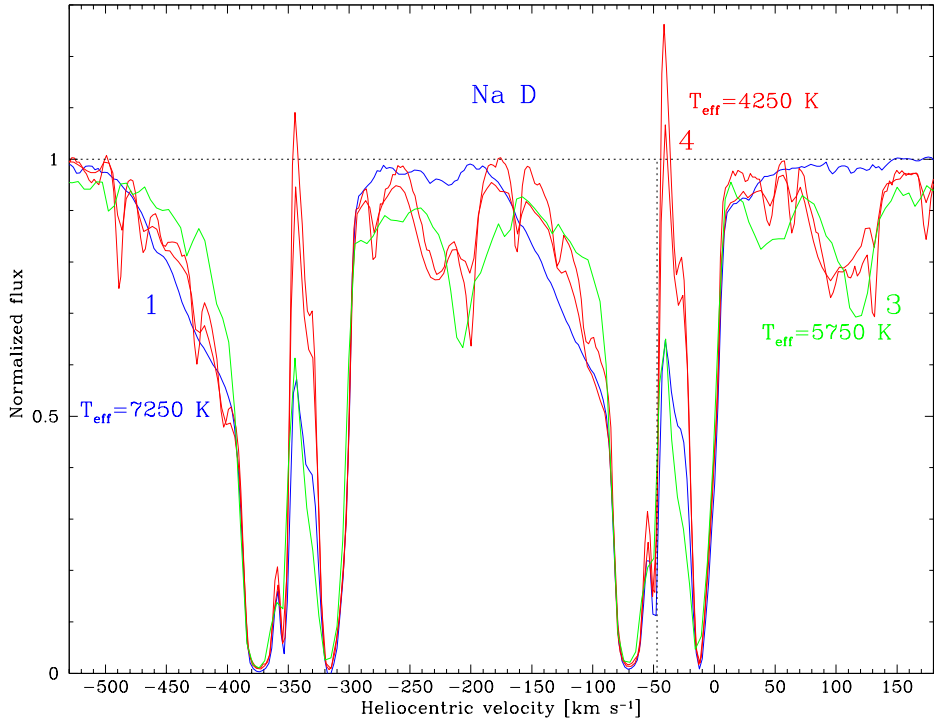}{5.0cm}{0}{50}{50}{-170}{-140}
\vspace*{4cm}
\caption {During the outburst central emission appears 
above the local continuum level in the Na~$D$ lines. The broad absorption portions of the lines are strongly intensity 
saturated, and do not show Doppler shifts with time. 
The maximum of the central emission line is red-shifted with respect to the stellar 
rest velocity, and a blue-shifted narrow feature is observed in both lines of the doublet.
} \label{fig-5}
\end{figure}

We compute from the synthetic spectrum fits to the TiO bands a 
mass-loss rate of $\dot{M}$$\simeq$5.4\,$\times$\,$10^{-2}$~$M_{\odot}\,yr^{-1}$ during the outburst 
of $\rho$ Cas. We assume spherical geometry; $\dot{M}$=4\,$\pi$\,$R_{\star}^{2}$\,$\rho$\,$v_{\rm exp}$, 
with $R_{\star}$=400$\pm$100~$R_{\odot}$, and $\rho$$\simeq$$10^{-10}$~$\rm gr\,cm^{-3}$ the minimum density 
in the TiO line formation region.
The spectrum synthesis reveals that $T_{\rm eff}$ returned to 
$\simeq$5750 K within 100 d after the deep outburst minimum ({\em magenta dotted line in 
Figure 3}), corresponding to a brightness increase of $\sim$$0^{\rm m}.5$ 
({\em magenta vertical line labeled 5 in Figure 1}). 

During the outburst, the H$\alpha$ line profile develops prominent emission wings, 
signaling a large $\dot{M}$, around a central absorption core 
that becomes unusually weak ({\it red line in Figure 4}). 
The comparison with the H$\alpha$ profiles of Betelgeuse and $\mu$ Cep
in Figure 4  ({\it lower black lines}) reveals that the line core during the outburst 
is not excited by a stellar chromosphere, as is the case for these M-type supergiants.
The spectrum computed for Betelgeuse without the model for the chromosphere 
({\em black dotted line}) shows that the H$\alpha$ transition is not sufficiently 
excited to match the depth of the observed absorption line core. The H$\alpha$
line profile of $\rho$ Cas is filled in by recombination line emission during the outburst cooling
of the upper atmosphere. The line thereby assumes a self-absorbed shape, indicating a very extended H$\alpha$ 
envelope which is optically thick.  
In 1999 May we observe ({\it green vertical line labeled 3 in Figure 1}) 
a very prominent blue emission wing in H$\alpha$, while the photospheric
spectrum strongly shifts to the longer wavelengths ({\it green line in Figure 4}). 
It indicates an exceptional (pre-outburst) variability phase with very strong downflow in the upper 
and lower atmosphere, which precedes the brightness maximum of 2000 March, leading up the 
actual outburst event. 

\section{Na~$D$ Emission During the Outburst}

During the outburst we observe prominent emission lines in the deep and broad  
absorption cores of the split Na $D$ doublet lines.
The central emission maxima appear {\it above} the local continuum level 
in Figure 5 ({\it red lines labeled 4}). This results from the strong 
decrease of the optical continuum flux due to 
the decrease of $T_{\rm eff}$ by at least 3000 K.
We observe that the TiO bands develop {\it before} the deep brightness minimum is reached,
when the photosphere assumes the maximum expansion velocity in Figure 1 ({\it vertical red line labeled 4}).
The $V$-brightness minimum results from a strong decrease of the entire 
atmospheric temperature structure. Unlike the photospheric absorption lines, the Na $D$ lines 
are intensity saturated and do not reveal Doppler shifts with time ({\it compare with 1 and 3 in Figure 5}). 
The far violet extended Na~$D$ line wings form in an optically thick wind during the fast atmospheric expansion.
These extended line wings were very strong in 1993 December ({\it blue line labeled 1}), with maximum wind 
velocities up to $-$130~$\rm km\,s^{-1}$.  

\begin{figure}
\plotone {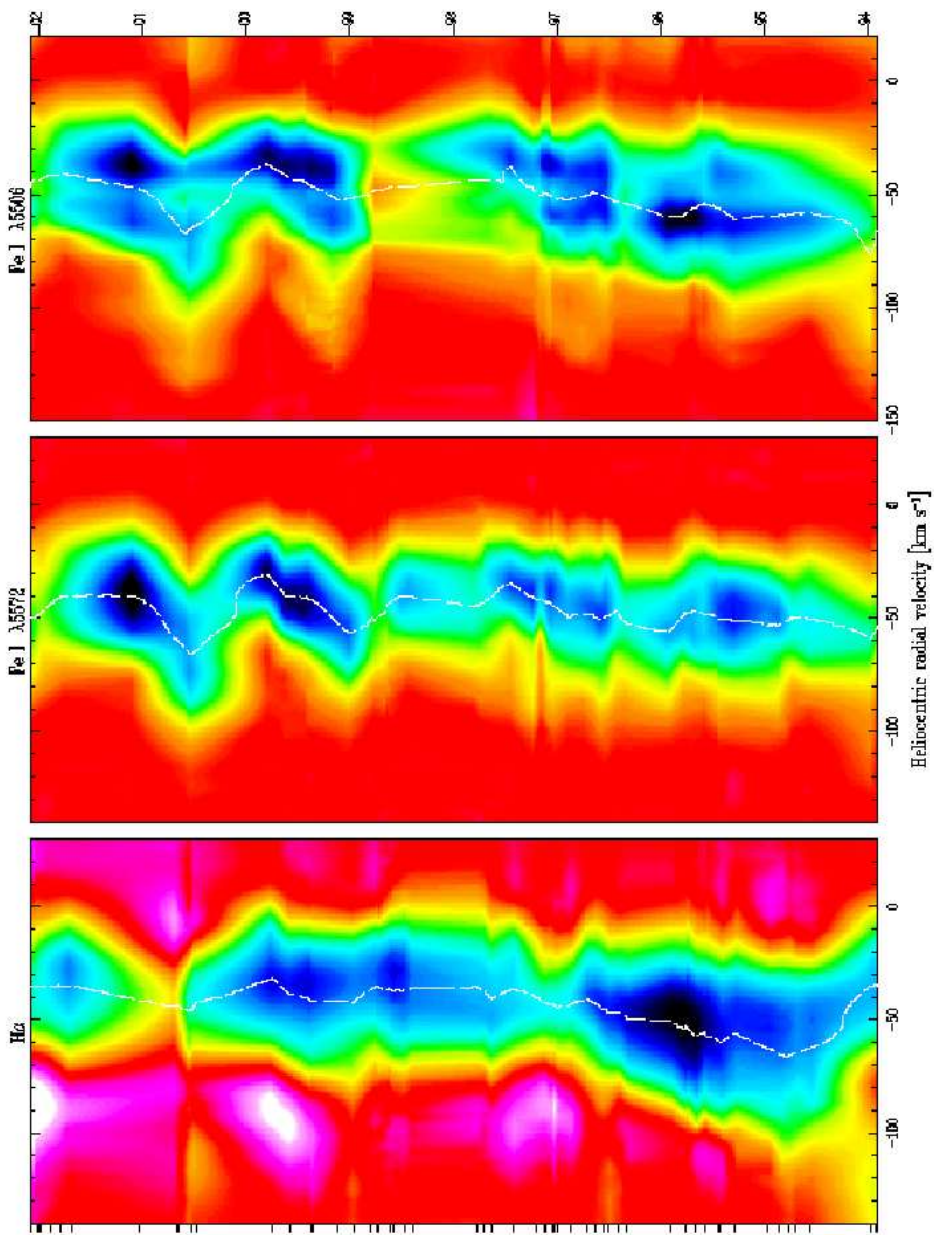} 
\caption {Dynamic spectra of H$\alpha$, Fe~{\sc i} $\lambda$5572, and the split Fe~{\sc i} $\lambda$5506 line. 
The white spots in H$\alpha$ are emission above the stellar continuum level. 
The line profiles are linearly interpolated between consecutive observation nights, marked
with the left-hand tickmarks. Time runs upward, indicated for each new calendar year with 
the right-hand numbers. The dashed white lines trace the radial 
velocity determined from the lines.
The curves reveal a strongly velocity-stratified dynamic atmosphere. 
Notice the strong blue-shift of the Fe~{\sc i} lines during the outburst (mid 2000). 
The outburst is preceded by very strong emission in the short-wavelength wing of H$\alpha$, 
while the absorption core extends longward, and the photospheric Fe~{\sc i} lines strongly red-shift. 
A strong collapse of the entire atmosphere precedes the outburst.
} \label{fig-6}
\end{figure}

\section{Dynamic Spectra of 1993--2002}

Figure 6 shows the dynamic spectra of H$\alpha$, Fe~{\sc i} $\lambda$5572, and the split Fe~{\sc i}
$\lambda$5506 line, observed between 1993 November and 2002 February. 
Red color indicates the stellar continuum level, and blue color the flux depression in the absorption lines. 
The white spots in H$\alpha$ are emission above the stellar continuum level. 
The line profiles are linearly interpolated between consecutive observation nights, marked
with the left-hand tickmarks.
The radial velocity curves at half intensity minimum of H$\alpha$, and of the Fe~{\sc i} lines 
({\it white dashed lines}) reveal a strongly velocity stratified dynamic atmosphere. We observe that 
H$\alpha$ varies over a much longer period of time compared to the photospheric Fe~{\sc i} lines.
It signals that H$\alpha$ forms on average higher, over a more extended portion of the pulsating 
atmosphere. Notice the strong blueshift of the Fe~{\sc i} lines during the outburst of mid 2000. 
The outburst is preceded by very strong line emission in the short wavelength wing of H$\alpha$, 
while the absorption core extends longward, and the photospheric Fe~{\sc i} lines strongly red-shift. 
We observed that a strong collapse of the entire atmosphere precedes the outburst event during 
the pre-outburst cycle of 1999. 

Notice further the development of even stronger emission in the short-wavelength wing of H$\alpha$, 
and very weak Fe~{\sc i} lines during the more recent brightness maximum of 2002 December.
High-resolution observations of 2002 June and July reveal a remarkable split H$\alpha$
absorption core. This type of line shape was not observed in the past 8.5 years.
Only Beardsley (1961) mentions the observation of ``core emission'' in H$\alpha$ 
during the famous outburst of 1945--47. Our H$\alpha$ observations may therefore signal that a 
new, presumably stronger, outburst is imminent in $\rho$ Cas.  

A spectral movie of the $\rho$ Cas outburst is electronically available at \\
{\tt http://cfa-www.harvard.edu/$\sim$alobel} through animated GIF and AVI format files.
A high-resolution atlas of the optical spectrum of $\rho$ Cas will also be posted at 
this URL address. 

\section{Hydrogen Recombination Driving Mechanism}

We compute that the hypergiant atmosphere becomes 
dynamically unstable because the first generalized adiabatic index $\Gamma_{1}$$\equiv$($d$\,ln\,$P$/$d$\,ln\,$\rho$$)_{\rm ad}$,
volume averaged over the atmosphere, assumes values below the stability value of 4/3
(Lobel 2001b). $<$$\Gamma_{1}$$>$ $<$4/3 when $T_{\rm eff}$ 
increases to $\geq$7250 K during the strong atmospheric collapse 
before the outburst. The outburst event, directly following this strong collapse, is driven by
the release of ionization energy due to the recombination 
of hydrogen with the cooling of the entire atmosphere by $\Delta$$T_{\rm eff}$$\simeq$3500~K.
For an adiabatic spherical expansion we calculate that 
the outburst time-scale is correctly predicted with (Lobel et al. 2003):

\begin{eqnarray}
t_{\rm burst} \simeq \frac{R_{\star}}{3\,\sqrt{N\,k\,T_{0}}}\, \frac{1}{\Gamma_{3}-1}\,  \frac{\frac{\Delta\,T_{\rm eff}}{T_{\rm eff}}}{\sqrt{\frac{5}{2}+5\,\frac{\Delta\, T_{\rm eff}}{T_{\rm eff}}+ \frac{I}{k\,T_{0}}}}\,  \\
\large
\noindent {\rm where} \,\,\,\frac{1}{\Gamma_{3}-1} = \frac{27 + 2\,(\frac{3}{2} + \frac{I}{k\,T_{0}})^{2}}{21+2\,\frac{I}{k\,T_{0}}}\,.  \nonumber 
\end{eqnarray}

$\Gamma_{3}$ is the third adiabatic index, evaluated for a mean ionization fraction of 1/2, in the partial hydrogen 
ionization zone, where the atmosphere is most expandable (or compressible).
For $R_{\star}$=400~$R_{\odot}$, $T_{\rm eff}$=7250~K, and $T_{0}$$\simeq$8000 K in the partial 
hydrogen ionization zone with $I$=13.6~eV, we compute $t_{\rm burst}$=221~d, and an
expansion velocity of 36~$\rm km\,s^{-1}$, in good agreement with the observed values.
Hence, equation (1) predicts that during the outburst the atmosphere expands over $\sim$2.5 $R_{\star}$ 
with the complete recombination of hydrogen.

We also estimate that the time-scale over which the 
total hydrogen recombination energy, released into the circumstellar environment by 
the outburst, is 
dissipated by the stellar radiation field during a period following the fast envelope expansion of 
\begin{equation}
t_{\rm brightness-min} \simeq \frac{I\,N\,{\rho}\, d_{\rm s}}{\sigma\, T_{\rm eff}^{4}} \,,
\end{equation}
where $\sigma$ is the Stefan-Boltzmann constant. For an atmospheric expansion of 
$d_{\rm s}$ = 869 $R_{\odot}$ during outburst, and a mean envelope density of
$\rho$ = $10^{-10}$ $\rm gr\,cm^{-3}$, 
we compute with equation (2) that $t_{\rm brightness-min}$$\simeq$86~d for $T_{\rm eff}$=3500 K. 
This time-scale corresponds very well to the period over which the deep brightness minimum of 
$\sim$100 d was observed in late 2000. The total energy released into the circumstellar 
environment $E_{\rm shell}$ = 4\,$\pi$\,$R_{\star}^{2}$\,$d_{\rm s}$\,$\rho$\,$N$\,$I$ with complete
hydrogen recombination is $\sim$$6.1\,\times$$10^{44}$~erg. This amount of energy yields
a luminosity of $L_{\rm shell}$ = 1.8\, $\times$\,$10^4$~$L_{\odot}$, when radiated during 100 d.   
For a total stellar luminosity of $L_{\star}$= 3.5 $\times$ $10^{5}$~$L_{\odot}$
during quiescent variability phases, $L_{\rm shell}$ is $\simeq$5\% of $L_{\star}$.

In conclusion, we find that during the outburst the partial hydrogen recombination zone
of the hypergiant atmosphere dissipates several percent of the stellar luminosity into the stellar 
environment, thereby driving an expanding wind to highly supersonic outflow velocities during a global 
cooling of the entire atmosphere by more than 3000~K.
\section{Conclusions}

We observe the formation of TiO bands in the spectrum of $\rho$ Cas before the deep brightness minimum
in the outburst of 2000-01. A supersonic expansion velocity is observed for the new TiO bands, while $V$ 
rapidly dims by more than a full magnitude. The TiO shell expands faster than the photosphere, observed from 
atomic metal absorption lines. A large oscillation cycle and a very bright light maximum, with $T_{\rm eff}$ above 7250 K,
precede the outburst event.
We compute that $\dot{M}$$\simeq$5.4\,$\times$\,$10^{-2}$ $M_{\odot}\,yr^{-1}$ 
during the outburst, whereby $T_{\rm eff}$ decreases from $\simeq$7250 to $<$3750 K. 
$T_{\rm eff}$ returns to 5750 K within 100 d after the deep outburst minimum.
Since recurrent outbursts occur about every half century in $\rho$ Cas, these outburst phases 
of punctuated mass-loss are the major mass-loss mechanism of this massive cool hypergiant.
A prominent emission line spectrum is observed above the stellar continuum level 
in $\rho$ Cas during phases of fast wind expansion. 
We compute that the outburst is driven by the release of hydrogen recombination energy with the 
cooling of the entire atmosphere.
Based on observations of the H$\alpha$ line profile evolution in the summer of 2002 
we expect that a new outburst of $\rho$ Cas is imminent.

\acknowledgments
This research is supported in part by an STScI grant GO-5409.02-93A to the 
Smithsonian Astrophysical Observatory. 
A. L. gratefully acknowledges support from SRON-Utrecht (The Netherlands) 
over the years for maintaining the spectral data base of $\rho$ Cas. We thank 
the numerous observers of the WHT-UES service programs who have contributed to this 
long-term project.

\end{document}